%% file: usenix.tex
\providecommand{\main}{.} 
\documentclass[sigconf]{acmart}

\usepackage{multirow}
\usepackage{tikz}
\usepackage{amsmath}
\DeclareMathOperator*{\argmax}{arg\,max}

\usepackage{enumitem}
\setlist[enumerate]{label=\textbf{\arabic*.}}
\usepackage{filecontents}
\usepackage{hyperref}
\usepackage{graphicx}
\graphicspath{{\main/images/}{images/}}
\usepackage{float}
\usepackage{subcaption}
\usepackage{mathtools}
\usepackage{breqn}
\usepackage{url}
\usepackage{listings}
\usepackage{xcolor}
\usepackage{xspace}
\usepackage{algorithm}
\usepackage[noend]{algpseudocode}
\usepackage{subfiles}

\usepackage[done=false, later=false, notes=true]{dtrt}

\colorlet{punct}{red!60!black}
\definecolor{background}{HTML}{EEEEEE}
\definecolor{delim}{RGB}{20,105,176}
\colorlet{numb}{magenta!60!black}

\algnewcommand\algorithmicforeach{\textbf{for each}}
\algdef{S}[FOR]{ForEach}[1]{\algorithmicforeach\ #1\ \algorithmicdo}
\lstdefinelanguage{json}{
    basicstyle=\normalfont\ttfamily,
    numbers=left,
    numberstyle=\scriptsize,
    stepnumber=0,
    numbersep=8pt,
    showstringspaces=false,
    breaklines=true,
    frame=lines,
    backgroundcolor=\color{background},
    literate=
     *{0}{{{\color{numb}0}}}{1}
      {1}{{{\color{numb}1}}}{1}
      {2}{{{\color{numb}2}}}{1}
      {3}{{{\color{numb}3}}}{1}
      {4}{{{\color{numb}4}}}{1}
      {5}{{{\color{numb}5}}}{1}
      {6}{{{\color{numb}6}}}{1}
      {7}{{{\color{numb}7}}}{1}
      {8}{{{\color{numb}8}}}{1}
      {9}{{{\color{numb}9}}}{1}
      {:}{{{\color{punct}{:}}}}{1}
      {,}{{{\color{punct}{,}}}}{1}
      {\{}{{{\color{delim}{\{}}}}{1}
      {\}}{{{\color{delim}{\}}}}}{1}
      {[}{{{\color{delim}{[}}}}{1}
      {]}{{{\color{delim}{]}}}}{1},
}

\newcommand{\oursys}{{\sc BigFoot}\xspace}
\newcommand{\parabe}[1]{\vspace{0.75ex}\noindent{\bf \em #1}\hspace*{.3em}}

\begin{document}


\title{\Large \bf BigFoot: Exploiting and Mitigating Leakage \\
in Encrypted Write-Ahead Logs}

\author{
{\rm Jialing Pei and Vitaly Shmatikov}\\
Cornell Tech
} 


\begin{abstract}

Modern databases and data-warehousing systems separate query processing
and durable storage.  Storage systems have idiosyncratic bugs and security
vulnerabilities, thus attacks that compromise only storage are a realistic
threat.  In this paper, we show that encryption alone is not sufficient to
protect databases from compromised storage.  Using MongoDB's WiredTiger as
a concrete example, we demonstrate that \emph{sizes} of encrypted writes
to a durable write-ahead log can reveal sensitive information about the
inputs and activities of MongoDB applications.  We then design, implement,
and evaluate \oursys, a WAL modification that mitigates size leakage.

\end{abstract}

%
%

\maketitle

\section{Introduction}
\label{sec:intro}

\subfile{sections/Intro}
\section{Background}
\label{sec:background}

\subfile{sections/Background}


\section{Threat model}
\label{sec:threat}

\subfile{sections/ThreatModel}


\section{Inferring Secrets of MongoDB Applications}
\label{sec:attack}

\subfile{sections/Attack}


\section{Mitigating WAL Leakage}
\label{sec:sysover}

\subfile{sections/newSystem}




\section{Evaluation}
\label{sec:eval}

\subfile{sections/newEval}


\section{Limitations}
\label{sec:limit}

\subfile{sections/Limitation}


\section{Related work}
\label{sec:relatedwork}

\subfile{sections/newRelated}


\section{Conclusions}

Storage-only attacks are a realistic threat to modern databases and
data warehousing systems.  We demonstrated that encryption alone is
not enough to hide applications' secrets from compromised storage.
Fine-grained encrypted writes to the write-ahead log are correlated
with applications' inputs and activities and can thus reveal them to
a storage attacker.  We proposed \oursys, a segmentation and padding
scheme that equalizes the sizes of WAL writes.  We implemented \oursys
as a modification to MongoDB's WiredTiger storage engine and showed that
it mitigates inference attacks at a modest performance cost.

\parabe{Acknowledgments.}
Supported by NSF grant 1704296.  Thanks to Paul Grubbs for advice.

\bibliographystyle{plain}
\bibliography{cdb}

\end{document}

%% file: sections/Intro.tex
To take advantage of the elasticity of compute and storage resources
in the cloud, many modern databases and data warehousing systems
separate query processing and storage~\cite{verbitski2017amazon,
dageville2016snowflake, RedShiftSpectrum}.

\begin{figure}[h]
\centering
\includegraphics[width=0.4\textwidth]{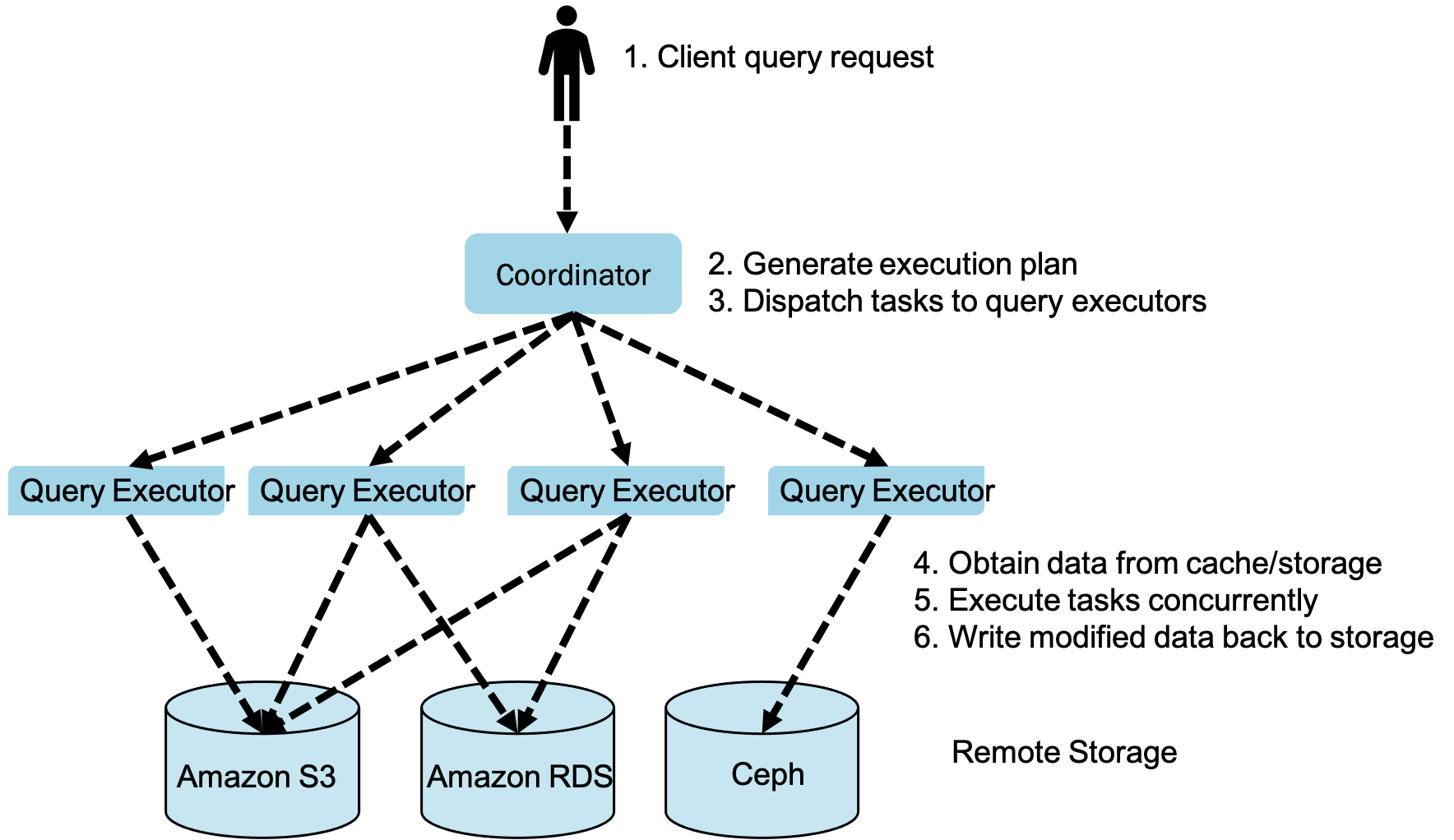}
\caption{Separation between query processing and storage.}
\label{fig:separation}
\end{figure}

The two components are independently managed and connected over a network
as shown in Figure~\ref{fig:separation}.  A coordinator receives clients'
requests and dispatches execution tasks to different query executors.
Executors retrieve the data from either local memory or remote
storage (e.g., a storage service such as Ceph~\cite{weil2006ceph} or
S3~\cite{s3}), process queries, and write modified data back to remote
storage for durability.  This helps reduce operational costs since
compute and storage can be provisioned and scaled independently.



Even databases that do not rely on cloud storage may use network-attached
storage (NAS) in lieu of a disk attached to the query server.  Persistent
storage systems, whether cloud storage or NAS, have their own software
stacks, including firmware, file system, OS, etc., distinct from the
servers that process queries.  Storage servers have their own bugs and
security vulnerabilities.  Consequently, storage should be viewed as
a separate trust domain.  Critically, \emph{persistent storage can be
compromised independently of the query processor}.  A recent example of an
attack that compromised storage without compromising query or application
servers is the 2019 Capital One breach, where a former Amazon Web Services
(AWS) engineer exploited a misconfiguration to obtain Capital One's AWS
credentials and used these credentials to gain unauthorized access to
106 million accounts~\cite{caponehack}.

Databases and data warehousing systems that disaggregate compute and
storage should protect the secrets of the applications that use them
(both data and queries) even if persistent storage is compromised.
Encrypting data in persistent storage, i.e., ``encryption-at-rest,'' is
not enough.  Encryption hides the content but not the sizes of storage
operations.  Variable-sized reads and writes whose sizes are correlated
with applications' activities can reveal sensitive information, such
as the values of inputs into these applications and the operations they
are performing.

\parabe{Our contributions.} 
First, we identify write-ahead logging (WAL), a standard durability
and concurrency control mechanism, as a source of fine-grained leaks
about applications' activities to compromised storage.  We use MongoDB's
WiredTiger storage engine to demonstrate how an attacker in control of
storage can infer secrets of a MongoDB-based application by analyzing
the sizes of encrypted WAL writes that result from the application's
activities.

Second, we design, implement, and evaluate \oursys, a segmentation and
padding scheme that ensures all writes to the persistent WAL have equal
sizes, thus mitigating the leakage.  We benchmark \oursys on concurrent
and sequential workloads, and measure the tradeoff between segment
size and performance overhead.  Small segments result in higher runtime
overhead for concurrent workloads (up to 7\% decrease in throughput and
13\% decrease in latency) but smaller space overhead for sequential
workloads, while large segments exhibit very similar latency and
throughput to unmodified WiredTiger at the cost of higher space overhead
(between 8x and 80x, depending on the distribution of write sizes).

%% file: sections/Background.tex
\subsection{Write-ahead logging}

Even databases that attempt to perform most operations in volatile memory
must achieve durability.  A popular technique is to \emph{checkpoint}
by flushing modified memory pages to disk at fixed time intervals.
Crashes between checkpoints, however, may result in a loss of writes
that have not yet been written to disk.  Write-ahead logging (WAL)
provides a way to make these writes durable~\cite{wal}.

In general, WAL is a family of techniques for providing atomicity and
durability (two of the ACID properties) in database systems.  WAL wraps
information about the current write and stores it durably before the write
is confirmed to the client application.  Usually a log sequence number
(LSN) is associated with each logged write to establish the happen-before
relation~\cite{lamport2019time} between logs.  Log records are stored
in a memory log buffer and later synchronously written to non-volatile
storage by the WAL protocol.  Upon failure, a data recovery scheme
like ARIES \cite{mohan1992aries} replays all logs in the LSN order to
reconstruct the state of the database immediately prior to the crash.

\subsection{Durability and recovery in MongoDB}

We use MongoDB~\cite{Mongo} as our case study and as the platform
for our prototype of \oursys.  MongoDB is a popular, open-sourced,
document-oriented database.  WiredTiger is Mongo's storage engine.

\parabe{Checkpoints.}
WiredTiger uses B-trees to store data in volatile memory.  A snapshot
is a consistent, durable view of these B-trees, written out to disk.
Starting from version 3.6, MongoDB configures WiredTiger to create
checkpoints (i.e., write the snapshot data to disk) every 60 seconds.

\begin{figure}[h]
\centering
\includegraphics[width=0.15\textwidth]{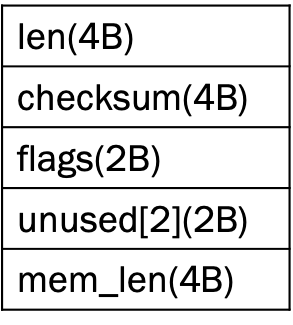}
\caption{Header structure in log records.}
\label{fig:header}
\end{figure}

\parabe{Write-ahead logging.} 
To provide durability in the event of a failure between checkpoints,
WiredTiger uses write-ahead logging to on-disk \emph{journal}
files~\cite{Mongojournal}.  WiredTiger creates one log record for each
client-initiated write operation.  A log record wraps all internal
write operations to the WiredTiger's in-memory data structures caused
by the application-initiated write.  A log record consists of a 16B
header and data.  Figure~\ref{fig:header} shows the layout of headers.
Note that (a) the first 4 bytes of the header of any log record are
not all zeros because they contain the length of the record, and (b)
this length is a multiple of 4 bytes.

MongoDB configures WiredTiger to buffer all log records up to 128 KB in an
in-memory data structure called the \emph{slot}.  Slots are synchronously
flushed to non-volatile storage every 100 milliseconds or upon a full-sync
write, whichever comes first.  A full-sync write is a write operation
that requires its journal record to be flushed to non-volatile storage
before returning, thus ensuring that the written data survives a crash.
A full-sync write provides the strictest durability, in contrast to
non-sync writes where the data is recorded in a buffer in memory but there
is no guarantee that it is immediately written to non-volatile storage.
After a full-sync write is issued by the client to the query executor,
all records in the slot buffer must be synchronized to non-volatile
storage in order to commit the write.

\begin{figure}[h]
\centering
\includegraphics[width=0.4\textwidth]{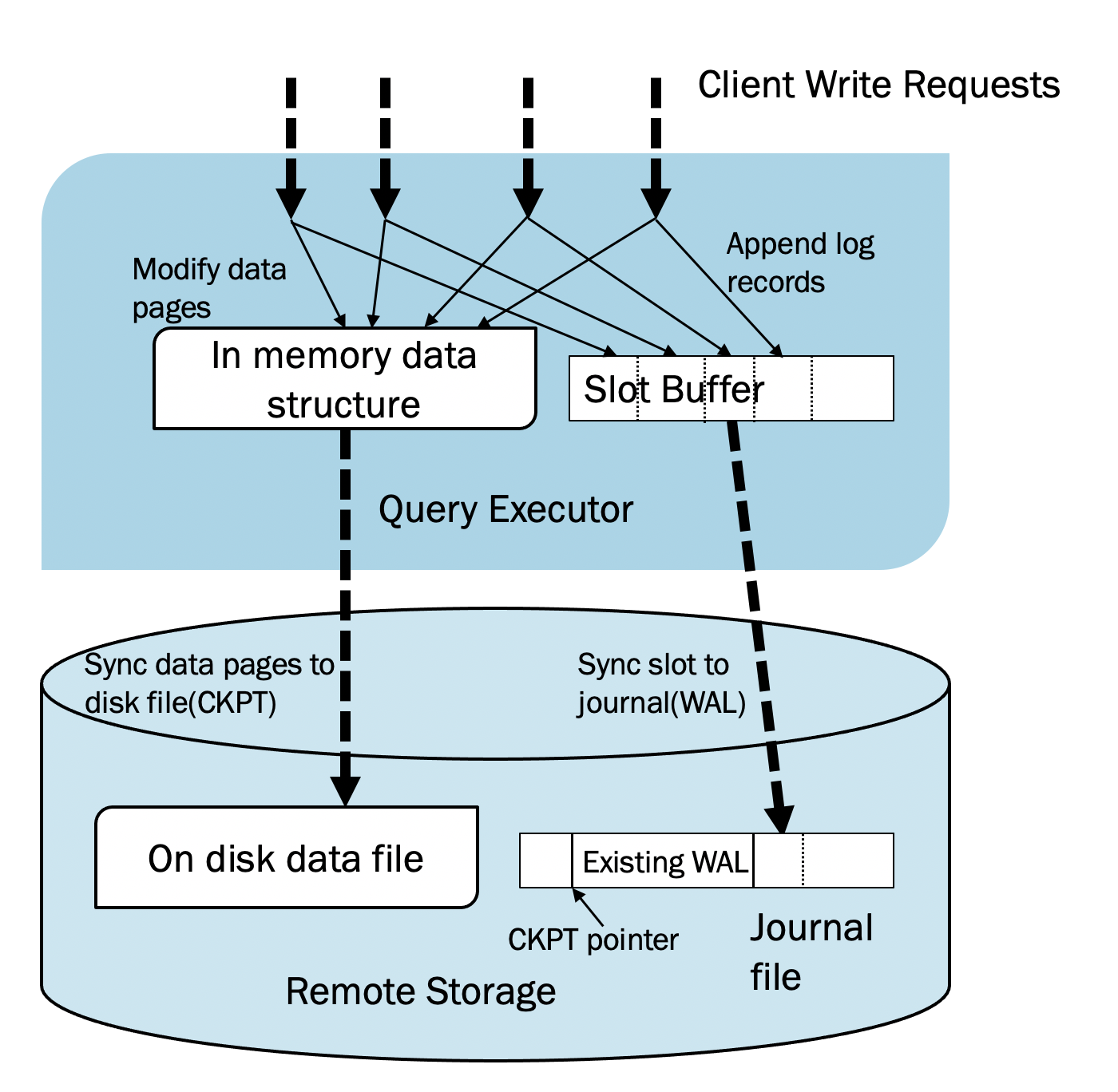}
\caption{Checkpoints and WAL in WiredTiger.}
\label{fig:mongockpt}
\end{figure}

Figure~\ref{fig:mongockpt} shows an overview of how MongoDB achieves
durability by periodically checkpointing memory data pages and appending
slots of journal records to a disk file between checkpoints.  After every
checkpoint, all journal records whose writes ``happened before'' the
checkpoint are automatically garbage-collected, freeing space for the
future journal records.  In this sense, the journal can be thought of
as a circular buffer for write operations.  WiredTiger keeps track of
the current checkpoint and the current starting point for WAL in the
journal file.

\parabe{Recovery.} 
When recovering from a crash, WiredTiger first looks in the data files for
the identifier of the last checkpoint, then searches the journal files
for the record that matches this identifier.  WiredTiger then reads log
records one by one from the journal file: First it scans through a 16B fixed size header to obtain meta information such as how large the whole record is. Second it reads through the data part of log record according to meta information.  After reading each record,
it immediately applies it and continues to the next record until all
records are consumed.

\parabe{Sharding.}
MongoDB uses sharding~\cite{Mongoshard} to support deployments with very
large datasets and high-throughput operations that might exhaust the
capacity of a single server.  In the sharding mode, data is partitioned at
the collection level.  A collection is shared by all nodes in the cluster;
each node runs a separate mongod~\cite{Mongod} instance and stores a
fraction of the collection.  Each node maintains its own WAL/journal,
thus a client-initiated write operation generates a journal record in
the node that holds the corresponding data.



%% file: sections/ThreatModel.tex
We consider an adversary who completely controls the persistent storage
but not the query processor (see Section~\ref{sec:intro}).  This adversary
observes all durable reads and writes performed by the processor.
Even if ``at rest'' encryption is deployed to hide the content, the
adversary still observes the sizes of all reads and writes.

\begin{figure}[h]
\centering
\includegraphics[width=0.4\textwidth]{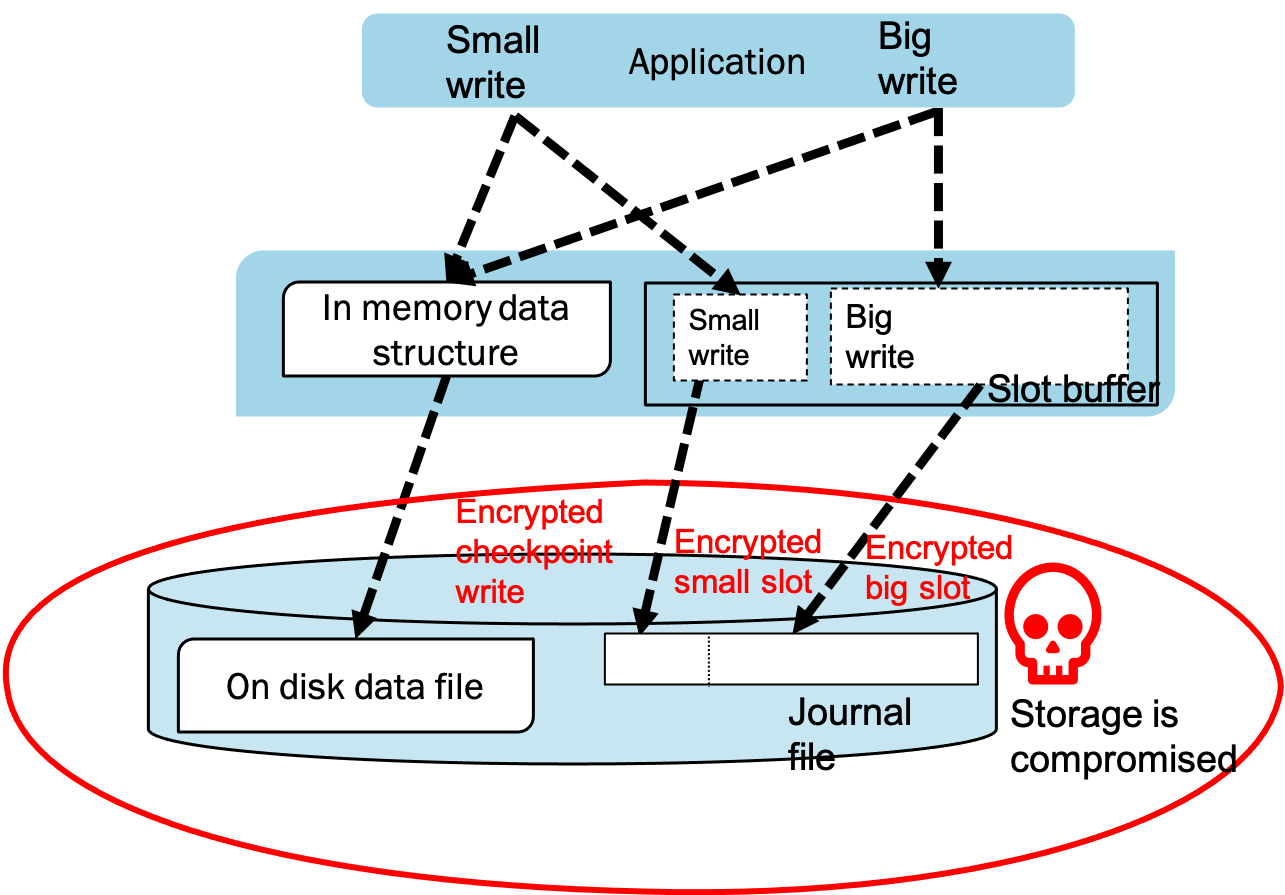}
\caption{Coarse- and fine-grained leakage in WiredTiger.}
\label{fig:leak}
\end{figure}

\parabe{Leakage in WiredTiger.}
For concreteness, Figure~\ref{fig:leak} shows what an adversary in
control of the disk (or any other persistent storage) observes in the
case of MongoDB's WiredTiger.  We assume that the application running on
top of MongoDB requires strong durability, thus every write operation
is issued as a full-sync write and the corresponding WAL increment is
immediately flushed to disk.

There are two types of writes in WiredTiger that can be observed by the
storage adversary: checkpoint writes and WAL slot writes.  Both writes
are encrypted but, as we show in Section~\ref{sec:attack}, their sizes
reveal information about the state of the application and even the data
it is operating on.

A checkpoint write happens at fixed time intervals and commits to disk
the result of multiple writes to the in-memory data structure.  This is
a coarse-grained leak.  To achieve durability between checkpoints, a WAL
slot write follows every application-initiated write to the in-memory
slot buffer.  If the application write is small (respectively, large),
the corresponding incremental write to the on-disk journal is small
(respectively, large).  This is a fine-grained leak.  


Another potential source of leakage is database \emph{logs} which, in the
case of MongoDB, are stored on disk in plaintext~\cite{MongoLogging}.
Depending on the configuration parameters controlling verbosity,
logs may contain sensitive information about the MongoDB instance and
applications running on top of it.  The enterprise version of MongoDB uses
log redaction to suppress messages associated with logged events, leaving
only metadata, source file names, or line numbers related to the event.
For example, if the logging message describes the tuple being inserted,
log redaction modifies the log to remove the content of the tuple.


\parabe{Out-of-scope threats.}
As explained above, we assume that the query processor is separate
from storage.  For the purposes of this paper, all attacks on the query
processor are out of scope.  In particular, we assume that the attacker
does not have access to the query processor's memory.

In the rest of this paper, we focus primarily on the encrypted,
incremental WAL writes because their sizes reveal the most fine-grained
information about applications' activities.  Leakage from checkpoint sizes
may reveal coarse-grained information (such as relative changes in the
overall memory footprint), but we do not investigate it in this paper.
Attacks that exploit the relative timing of writes (e.g., how quickly
writes follow each other) are also out of scope for the purposes of
this paper.

%% file: sections/Attack.tex
In this section, we use MongoDB's WiredTiger storage engine to demonstrate
how the sizes of encrypted, incremental writes to the journal file can
reveal applications' secrets.  We focus on two types of secrets: inputs
into an application and operations performed by an application.

For the purposes of this section, we assume that the storage adversary
knows which application is running on top of MongoDB.  This information
may be available from the logs and/or other leakage channels unrelated
to the durability mechanisms.

\subsection{Inputs into an e-commerce application}

\begin{figure}[h]
\centering
\includegraphics[width=0.4\textwidth]{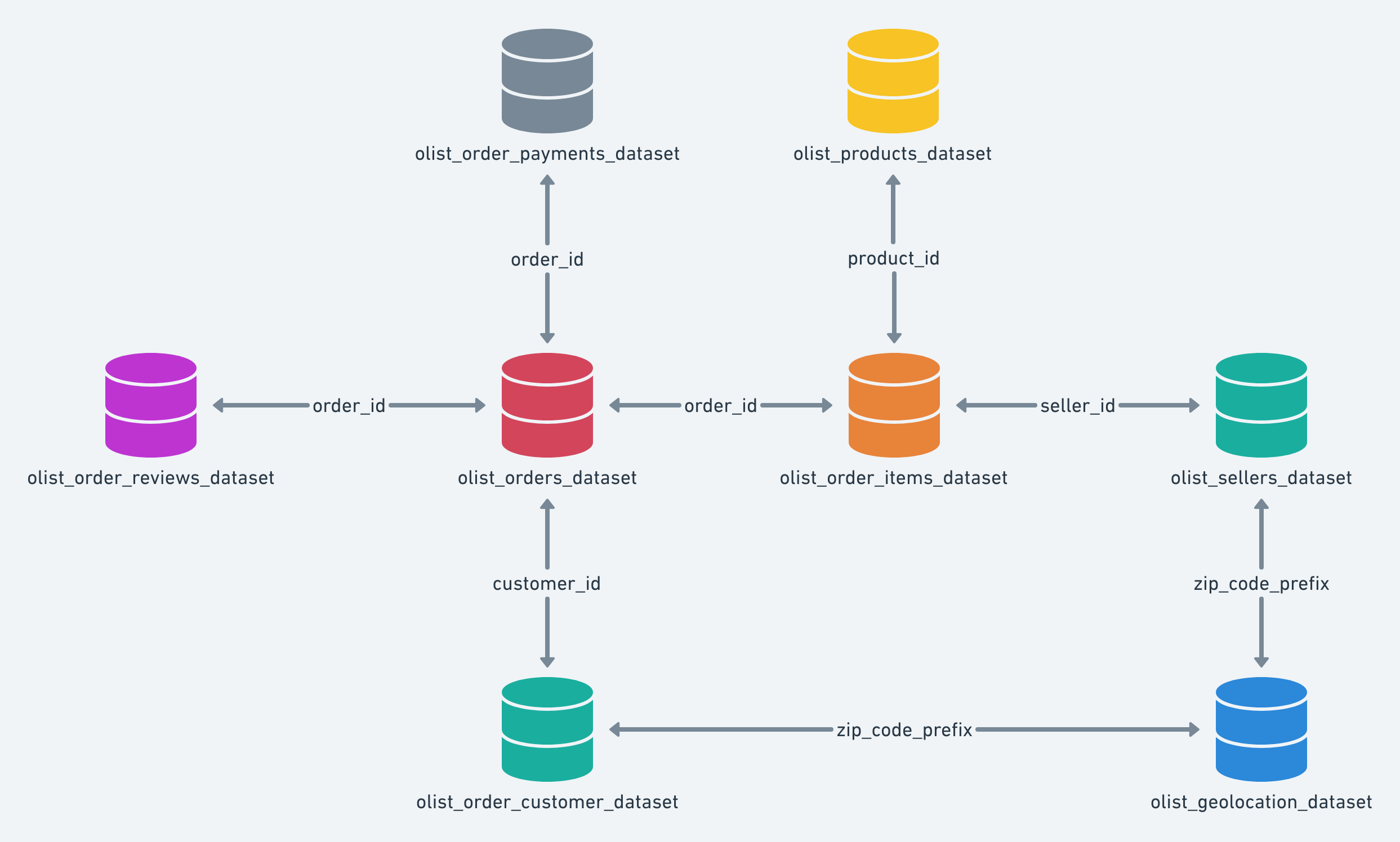}
\caption{Brazilian E-Commerce Public Dataset Schema.}
\label{fig:brazDS}
\end{figure}

We simulate a very simple e-commerce application using
the \emph{Brazilian E-Commerce Public Dataset} posted on
Kaggle.\footnote{https://www.kaggle.com/olistbr/brazilian-ecommerce} This
dataset has information about 100K orders made at multiple marketplaces
in Brazil from 2016 to 2018.  Figure~\ref{fig:brazDS} show the schema.
Information about each order includes order status, price, payment,
freight performance, customer location, product attributes, and customers
reviews.  We treat each table in the schema as a collection of documents
in MongoDB.

In our simulated workload, the application simply inserts records
from the olist\_customers\_dataset into MongoDB one by one.  The
attributes of these records are customer\_id, customer\_unique\_id,
customer\_zip\_code\_prefix, customer\_city, and customer\_state.

\parabe{Adversary's objective.}
Each record insertion results in an incremental write to the WiredTiger's
journal file.  Due to encryption at rest, the storage adversary observes
only the sizes of these writes.  The only source of variation in the
size of the record is the ``customer's city'' attribute.  The adversary's
goal in this case is to use the size of the WAL write to infer the size
of the original record written by the application and, from that, the
value of the ``customer's city'' attribute.  This information can have
commercial value.  For example, it can help estimate where a company is
expanding or seeing the biggest growth.

\begin{figure}[h]
\centering
\includegraphics[width=0.4\textwidth]{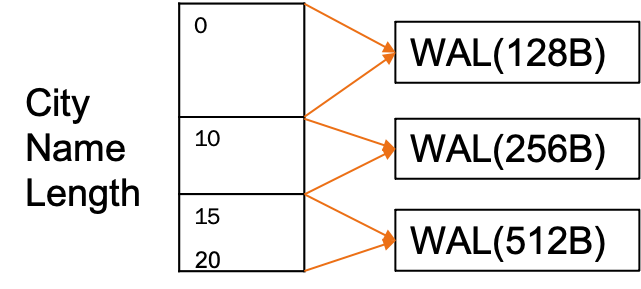}
\caption{Records of different sizes may produce equally-sized WAL writes.}
\label{fig:walmap}
\end{figure}

Directly inferring the value of the ``customer's city'' attribute is
not feasible for two reasons: (1) some city names have the same length,
thus the corresponding records have the same size, and (2) there are
only six distinct sizes of WAL writes that can be produced by records
from the olist\_customers\_dataset because WAL write sizes are always
multiples of 128 bytes.  Figure~\ref{fig:walmap} illustrates the latter:
records with city names whose length is under 20 produce WAL writes of
only 3 different sizes.

Therefore, our adversary has a coarser objective: given the size of a WAL
write, infer the \emph{range} of potential city names in the original
record.  This still leaks information because many ranges contain only
a few city names, greatly reducing the number of possibilities.

\parabe{Inferring the range of city-name lengths from WAL writes.}
We split the olist\_customers\_dataset into the training data (81511 rows)
and test data (17930 rows).  We use the training data to train a simple
inference model as follows.  Let S be all possible mappings from the size
of a WAL write to the range of city-name lengths.  Define $pred(r,w,m)$
to be a function that takes a write $r$, the corresponding WAL write $w$,
and a mapping $m$, and returns $1$ if $r$'s real size lies within $m(w)$,
$0$ otherwise.  To find the best mapping $m$, we solve:
\begin{equation}
 \argmax_{m \in S}{\sum\limits_{r,w}pred(r,w,m)} 
 \label{equ:opt}
\end{equation}

When evaluating the resulting model on test data, we measure precision
(for a given range, the ratio between the correct predictions and the
total number of predictions) and recall (the ratio between the correct
predictions for a given range and the true total size of that range).
Table~\ref{table:Precision and Recall} shows accuracy achieved by the
optimal mapping we found.

\begin{table}[h]
\centering
\resizebox{\linewidth}{!}{
\begin{tabular}{|r|c|r|c|c|}
\hline
WAL Size (B) & City Name Range & \# of City Names & Precision & Recall \\ \hline
128 & (0, 15]  & 3270 & 0.90 & 0.87 \\ \hline
384 & (15, 18] & 362 & 0.31 & 1.00 \\ \hline
512 & (18, 21] & 293 & 1.00 & 0.18 \\ \hline
768 & (21, 24] & 148 & 0.49 & 1.00 \\ \hline
1152 & (24, 27] & 41 & 1.00 & 0.15 \\ \hline
1536 & (27, 30] & 5 & 1.00 & 0.59 \\ \hline
\end{tabular}}
\caption{Precision and recall for different WAL sizes.}
\label{table:Precision and Recall}
\end{table}

Table~\ref{table:Precision and Recall} shows the tradeoff between
the accuracy of inference and the coarseness of inferred information:
the bigger the interval, the higher the precision and recall.  Bigger
intervals contain more city names, thus the adversary infers less
information about the true value of the city name in the record that
generated a given WAL write.  Smaller intervals offer lower precision
\emph{or} lower recall, but the adversary infers more information.
Accuracy of inference is not uniform across intervals, e.g., longer city
names are more vulnerable.


\subsection{Resources of a medical application}

To illustrate another type of adversarial inference, we show that a
storage adversary can learn sensitive information by inferring the
\emph{operations} performed by medical applications running on MongoDB
(as opposed to inputs into these applications).

Fast Healthcare Interoperability Resources (FHIR)~\cite{FHIR} is the
next-generation standards framework for health-care data exchange.
FHIR-based data systems provide schemas and REST APIs specified in the
standard to enable data exchange over HTTP between medical applications.
FHIR-enabled applications include mobile and cloud-based health-care
apps, EHR systems, server communication in institutional health-care
providers, etc.  FHIR solutions are built from of modular components
called ``resources.''  A resource represents any health-care concept,
e.g., patient, appointment, or medication, in a JSON form consisting of
key-value pairs\textemdash see an example in Listing~\ref{lst:schedule}.

\begin{lstlisting}[caption={Schema for ``Schedule'' resource.},label={lst:schedule},language=json,firstnumber=1,captionpos=b]
{
  "resourceType" : "Schedule",
  "identifier" : [{ Identifier }], 
  "active" : <boolean>, 
  "serviceCategory" : [{ CodeableConcept }], 
  "serviceType" : [{ CodeableConcept }], 
  "specialty" : [{ CodeableConcept }], 
  "actor" : [{ Reference(Patient|Practitioner|PractitionerRole|RelatedPerson|
   Device|HealthcareService|Location) }], 
  "planningHorizon" : { Period }, 
  "comment" : "<string>" 
}
\end{lstlisting}

We simulate a very simple medical application that inserts instances of
FHIR resources as documents into MongoDB.

\parabe{Adversary's objective.}
As before, the storage adversary only observes the sizes of WAL writes.
In this case, the size of a WAL write is largely determined by the
number of key-value pairs in the resource written by the application.
The adversary's objective is to infer the type of this resource and thus
the operation performed by the application.

\begin{figure}[h]
\centering
\includegraphics[width=0.5\textwidth]{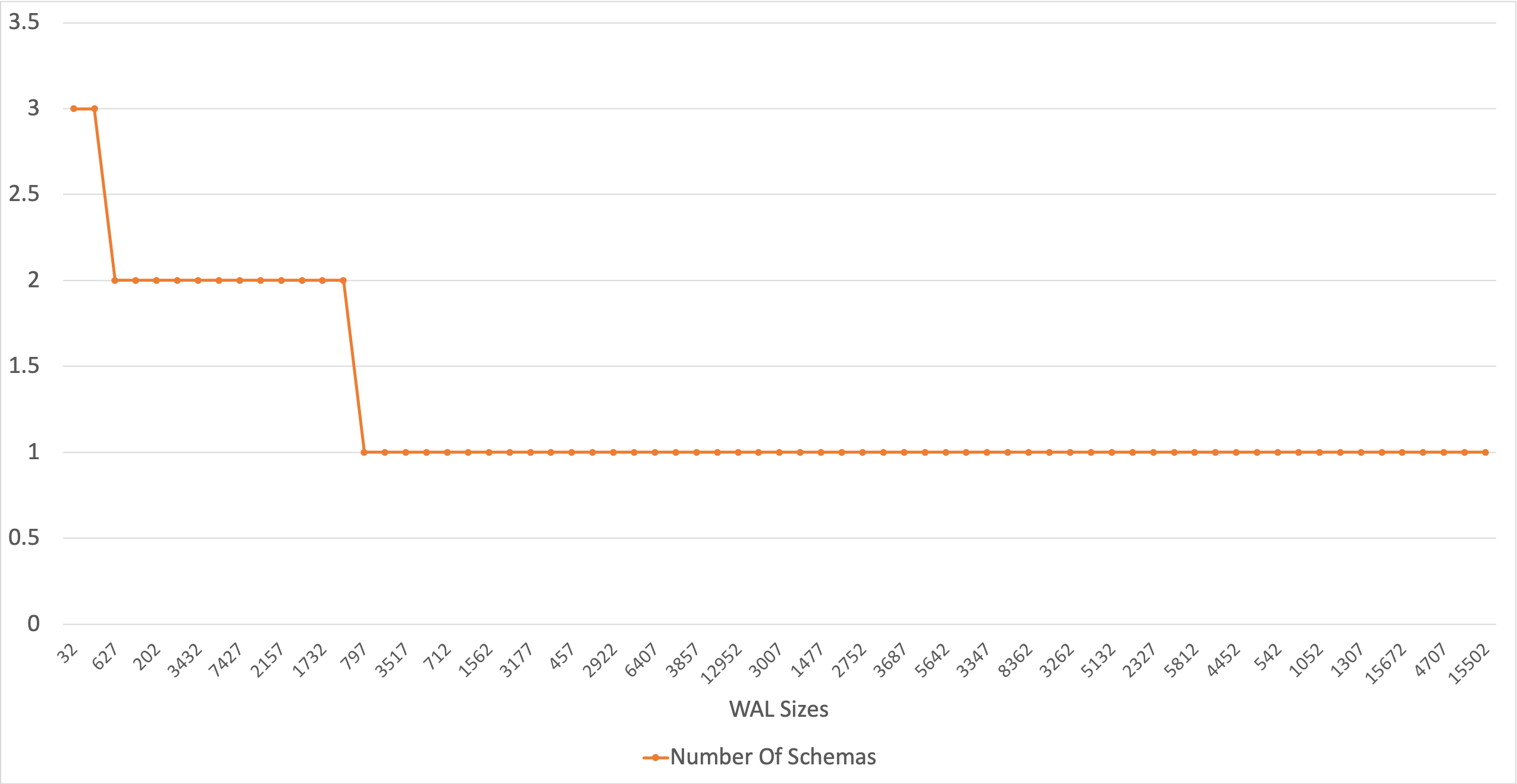}
\caption{Number of schemas corresponding to different WAL write sizes.}
\label{fig:wal2schema}
\end{figure}
\begin{figure}[h]
\centering
\includegraphics[width=0.5\textwidth]{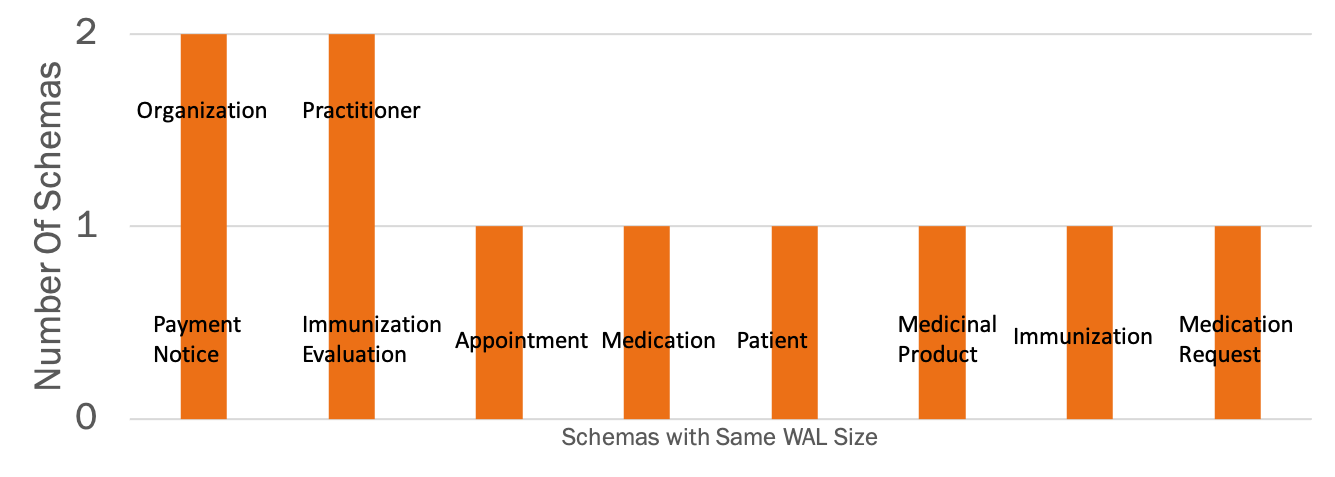}
\caption{Examples of schemas that can be inferred from WAL write sizes.}
\label{fig:zoominschema}
\end{figure}

\parabe{Inferring resource type.} 
In our simulated workload, we create 10 instances for each resource type
and randomly insert them one by one into the database in the full-sync
mode.  For each resource type, we record its schema and the average of
WAL write sizes resulting from the insertion of resources of this type.
We then construct a reverse mapping from WAL write sizes to resource
schemas, shown in Figure \ref{fig:wal2schema}.  Two WAL write sizes map
to 3 schemas each, twelve WAL write sizes map to 2 schemas each, and
fifty-five WAL write sizes map to unique schemas.  In the latter case,
the adversary observing a particular WAL write size can infer the schema
and resource type inserted by the application with 100\% confidence.
Figure~\ref{fig:zoominschema} shows examples of sensitive schemas that
can be inferred in this way.





%% file: sections/newSystem.tex
To mitigate leakage, we propose a modification to WiredTiger's WAL
mechanism.  Our modification, \oursys, ensures that \textbf{every WAL
write to persistent storage has the same size}.  This property ensures
that the storage attacker cannot exploit the sizes of WAL writes to
infer secrets of MongoDB applications.


\subsection{Segmenting WAL slots}
\label{segmentationoverview}

The naive solution to the problem of variable-sized slot writes is to
pad every write to the full slot size.  This solution is infeasible due
to prohibitive space overhead (see Section~\ref{sec:space}).

\begin{figure}[h]
\centering
\includegraphics[width=0.4\textwidth]{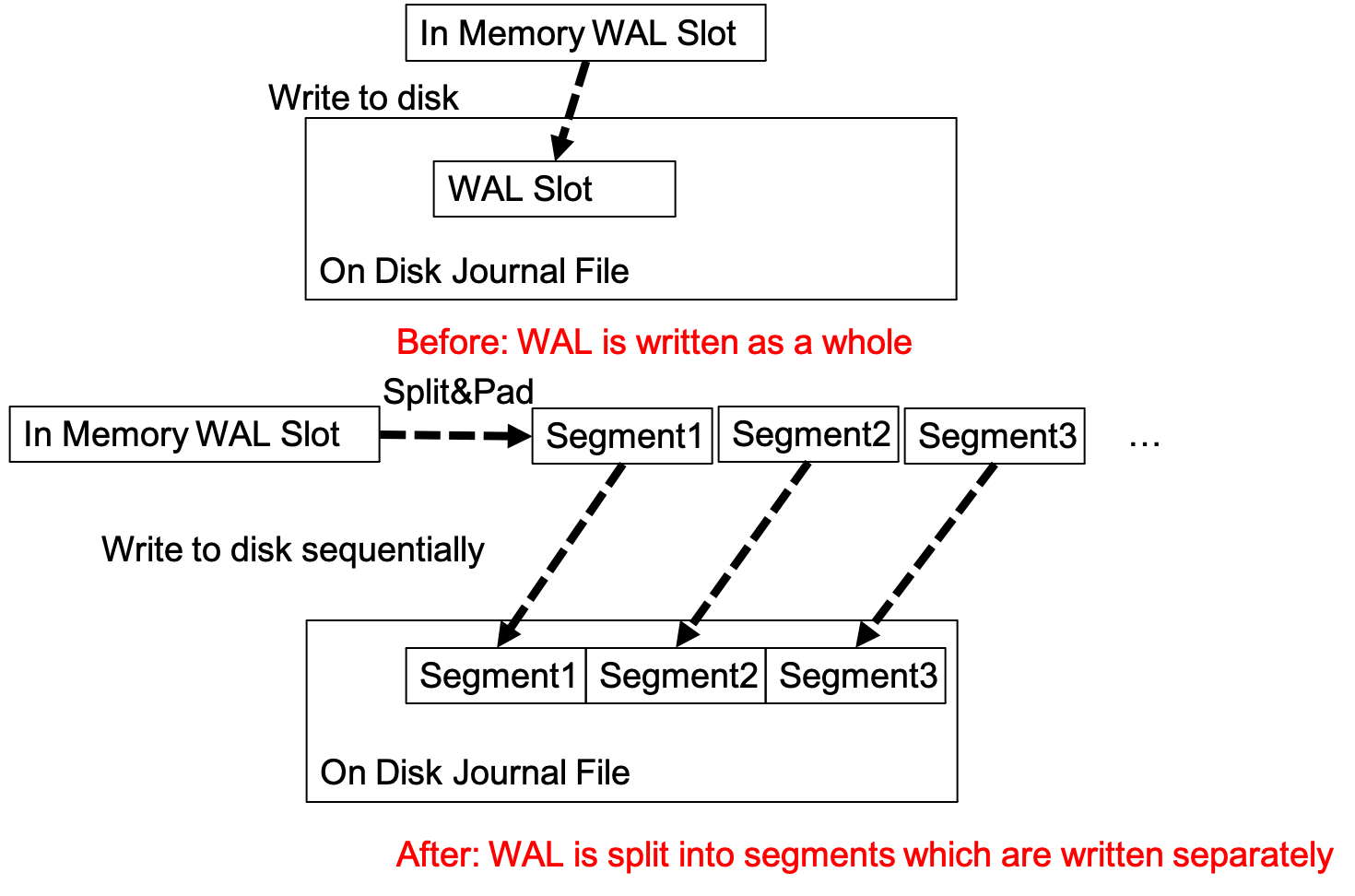}
\caption{Segmenting slot writes.}
\label{fig:segmentation}
\end{figure}

Figure~\ref{fig:segmentation} shows how \oursys works.  It splits
each slot write into fixed-size segments, pads the last segment with
$'0'$ if necessary, then writes each segment to the on-disk journal
file sequentially.  Because the lengths of log records and segments
are multiples of 4, the padding always consists of one or more 4-byte,
all-zero blocks.  Algorithm~\ref{segmentationalgo} shows the pseudocode
of the segmentation algorithm, parameterized by the segment size $S$.

\begin{algorithm}
    \caption{Split slot into segments}\label{segmentationalgo}
    \hspace*{\algorithmicindent} \textbf{Input: Slot SL, Segment Size S} \\
    \begin{algorithmic}[1]
    \Function{Segmentation}{SL,S}
        \State SSET = \{\};
        \While {SL.size() $\ge$ S}
        \State seg, SL = truncate(SL, S); // Split off a segment
        \State writeToJournal(seg);
        \EndWhile
        \State Pad remaining SL so that SL.size()==S;
        \State writeToJournal(SL);
    \EndFunction
    \end{algorithmic}
    \end{algorithm}

\label{subsec:segmentationchoice}

Segment size is the key configuration parameter.  Small segments reduce
the space overhead caused by padding writes to the segment boundary,
but they also increase the latency of big writes by increasing the
total number of segments (since segments are written separately and
sequentially).  Big segments, on the other hand, increase the space
overhead and reduce latency.  In Section~\ref{sec:eval}, we evaluate
this tradeoff.

\subsection{Recovery}

When recovering from a crash, \oursys reads from the last checkpoint in
the journal file, same as the unmodified WiredTiger.  The recovery process
works the same as in the unmodified WiredTiger whenever it encounters a
non-zero header.  After reading a full log record whose length is known
from the header, it skips all 4-byte, all-zero blocks (all such blocks
must be padding\textemdash see above) until it encounters a 4-byte,
non-zero block, which must be the beginning of the header of the next
log record.  The recovery process ends when all log records in the
journal file have been consumed.

%% file: sections/newEval.tex
Compared to the current WAL implementation in WiredTiger, \oursys imposes
a \emph{latency overhead} due to segmenting each slot write into multiple
segment writes and a \emph{space overhead} due to padding all writes
to the segment boundary.  The latency overhead is most severe for large
and/or concurrent writes when slots are mostly full.  The space overhead
is most severe for small, sequential writes when slots are mostly empty.
Large segment sizes incur lower latency overhead but higher space
overhead; small segment sizes have the opposite effect.

To evaluate the tradeoff between the segment size and these overheads,
we use (a) micro-benchmarks to measure the latency of single writes, and
(b) standard concurrent benchmarks to measure latency and throughput when
slots are full.  To measure space overhead, we use a sequential version
of the standard benchmark that processes and persists write queries one
by one.  We also compare recovery time with the current implementation.

\parabe{Experimental setup.}
We deployed a MongoDB instance in an AWS Virtual Private Cloud, configured
with 16 cores, 128GB DRAM and 256GB Elastic Block Store (EBS) and running
64-bit Ubuntu 16.04 with Linux kernel 3.2.0-23.

For our \emph{Single Query} micro-benchmark, we use insert queries
of different sizes: small (128B), medium (512B) and large (1KB).
Our macro-benchmarks are based on the Yahoo! Cloud Serving Benchmark
(YCSB), designed for performance comparisons of key-value storage systems.
The original YCSB contains Insert, Delete, Update, and Read operations.
Since \oursys changes only the processing of writes, we created an
all-insert version of YCSB.  Each write operation inserts a tuple
of 10 attributes; the size of each attribute is drawn from a certain
distribution.

We experimented with several distributions, including uniform, Zipf,
and constant.  For Zipf distributions, the $\alpha$ parameter controls
the skewness of the distribution.  For constant-size experiments,
we experimented with $S=200B$ and $S=800B$ as the write size.  In all
cases, the size of each attribute is smaller than or equal to 100B,
thus the maximum write size is 100*10=1000B.

As explained above, we use the concurrent version of this benchmark to
measure the latency and throughput overhead and the sequential version
to measure the space overhead.  \emph{Concurrent YCSB} processes
all writes concurrently.  For this benchmark, the full-sync mode is
disabled in MongoDB, therefore WAL writes are buffered in the slot and
the entire slot is flushed to disk periodically.  \emph{Sequential YCSB}
processes the same writes sequentially, with the full-sync mode enabled.
Therefore, each WAL write is flushed to disk individually.

\subsection{Latency overhead of a single write}

\begin{figure}[h]
\centering
\includegraphics[width=0.4\textwidth]{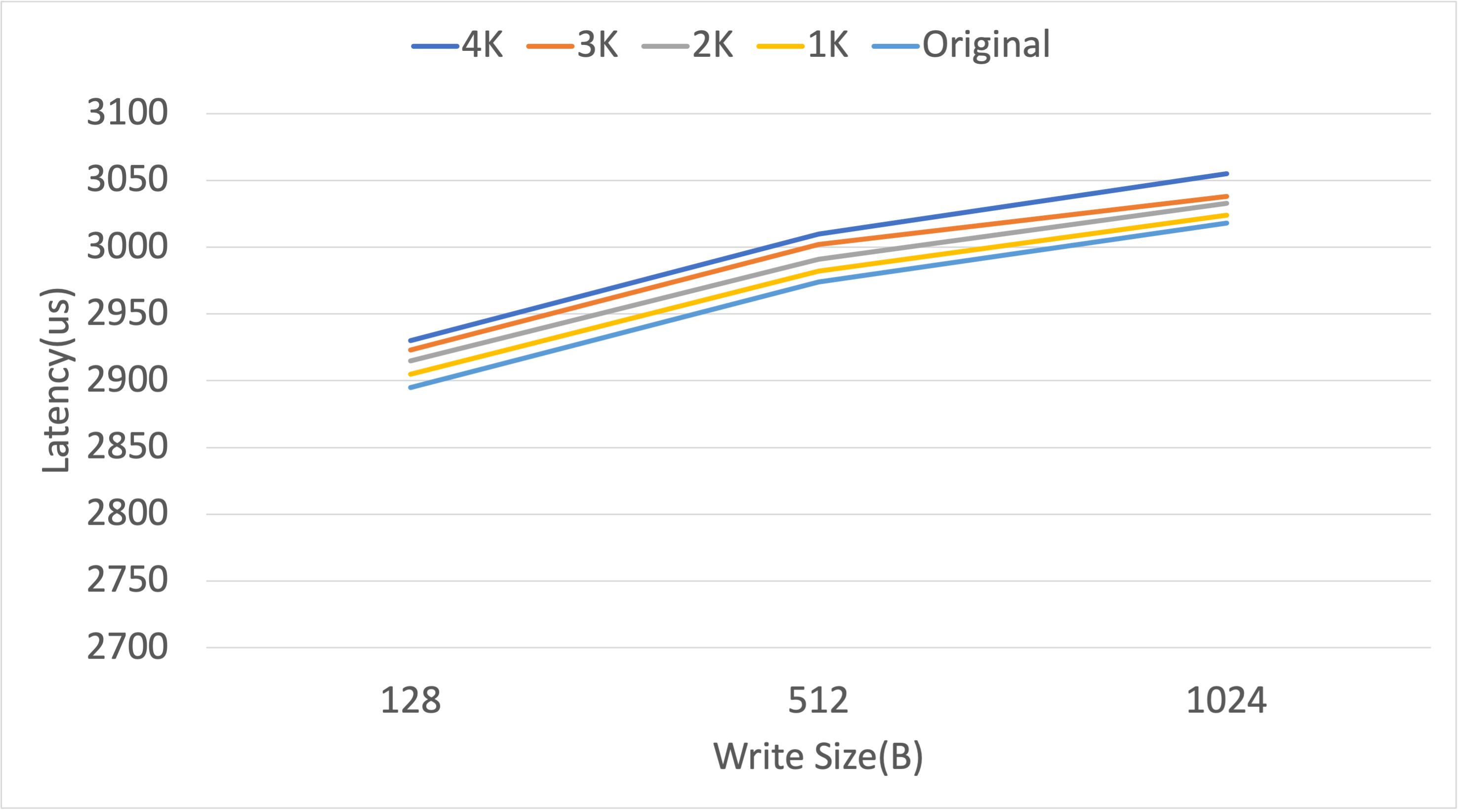}
\caption{Latency of a single write.}
\label{fig:latencySQ}
\end{figure}

Figure~\ref{fig:latencySQ} shows how latency of a single write varies
depending on the write size in the original WiredTiger and \oursys
with different segment sizes.  As expected, latency increases with the
segment size in all systems.  \oursys imposes $0.3-1\%$ extra latency
because padding increases the amount of data to be written, more so when
segments are large.

\subsection{Throughput and latency overhead of concurrent writes}

\begin{figure}[h]
\centering
\includegraphics[width=0.4\textwidth]{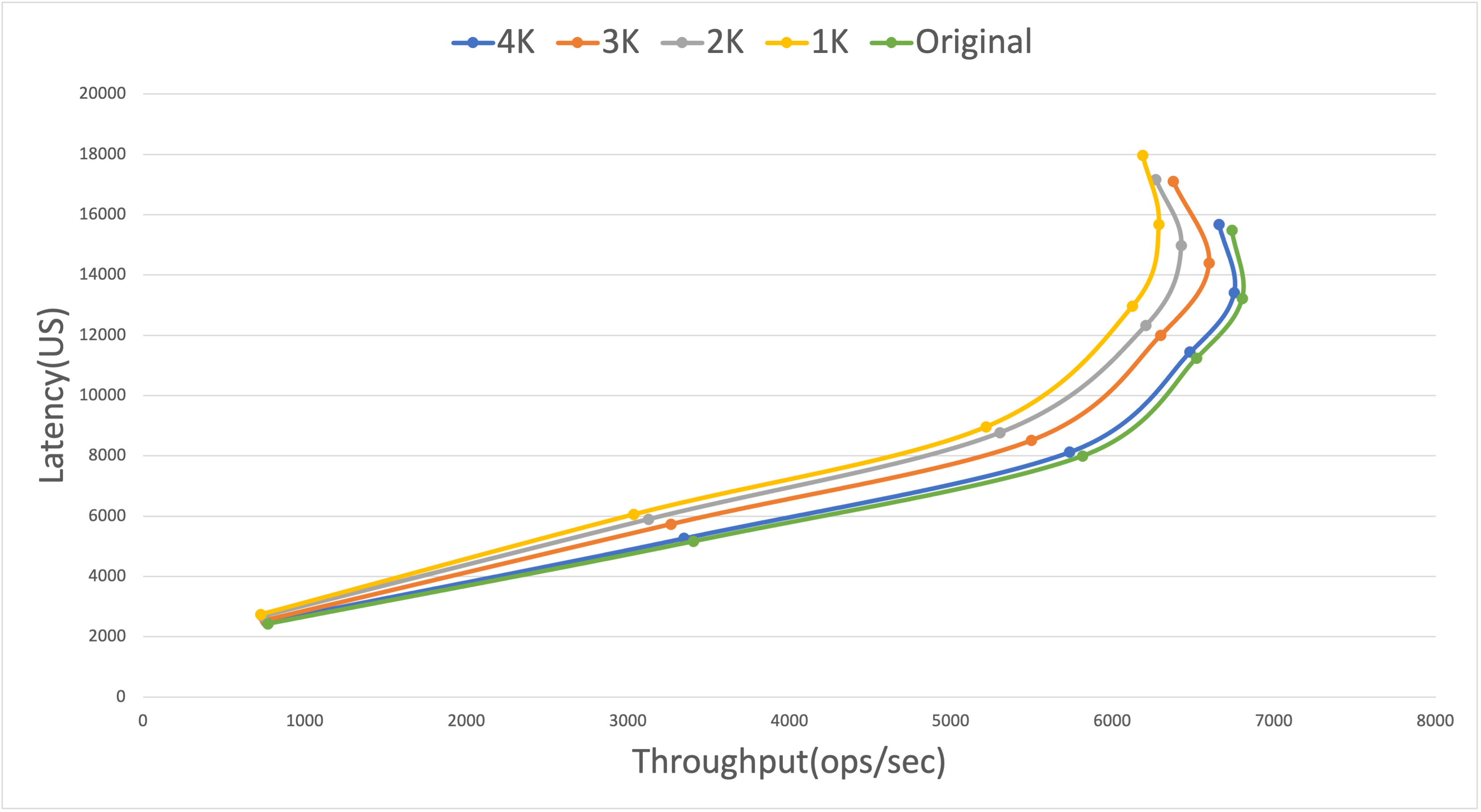}
\caption{Throughput and latency of concurrent writes.}
\label{fig:tpvsLat}
\end{figure}

Figure~\ref{fig:tpvsLat} shows throughput and latency under different
concurrency levels.  Writes for this experiment are generated from
a uniform (0,1K) distribution, following YCSB.  Latency increases as
contention between threads becomes more intensive.  Throughput starts
declining when the system is saturated.  The smaller the segments, the
earlier \oursys saturates (due to the increased number of writes for
each slot flush).  \oursys with 4K segments saturates at almost the same
point as the original WiredTiger.  Depending on the segment size, \oursys
incurs $1-7\%$ overhead in throughput and $3-13\%$ overhead in latency.

The distribution of write sizes does not have a significant effect
on the tradeoff shown in Fig.~\ref{fig:tpvsLat}.  Under the Zipf
and constant distributions, the absolute values of latencies and
throughputs, saturation points, and the differences between the original
WiredTiger and \oursys remain the same for all segment sizes as in
Figure~\ref{fig:tpvsLat}.

We conclude that \oursys imposes a modest performance overhead vs.\
the original WiredTiger in the worst scenario for \oursys, where all
all slots are full before they are flushed to disk.

\subsection{Space overhead}
\label{sec:space}

We measure the space overhead of \oursys by calculating Relative Cost (RC)
as the ratio between the amount of data written to disk by \oursys and
that written to disk by the original WiredTiger for the same queries.
For these experiments, we use the Sequential YCSB benchmark described
above.  This benchmark represents the worst-case scenario.  In the
original WiredTiger, it results in the biggest information leak to
the storage adversary due to the highly variable sizes of individual
WAL writes.  In \oursys, it results in the biggest space overhead due
to the padding of each write.

\begin{figure}[h]
    \centering
    \includegraphics[width=0.4\textwidth]{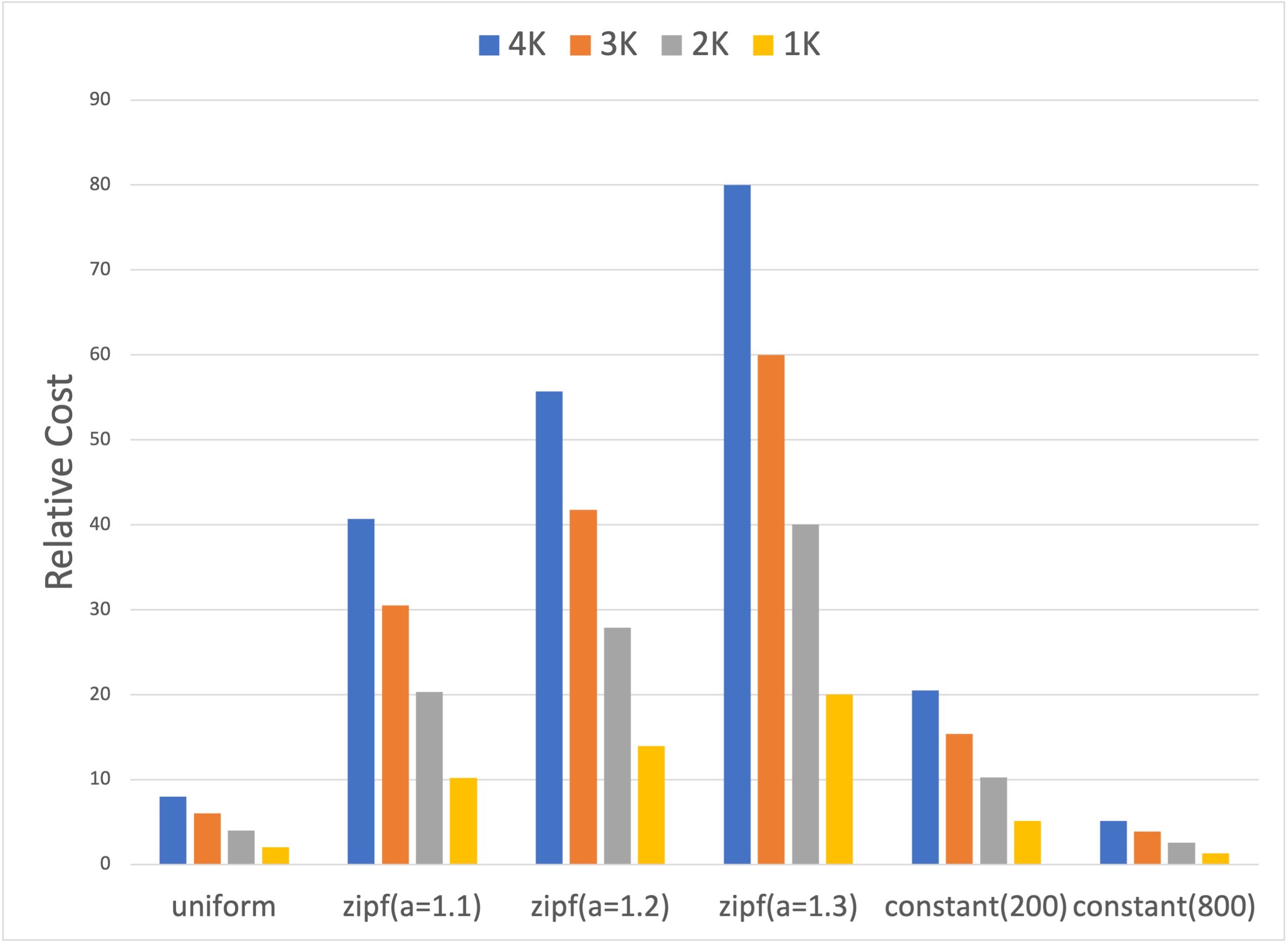}
    \caption{Relative space overhead for different distributions.}
    \label{fig:relativeCost}
\end{figure}

Under the uniform write-size distribution in the YCSB benchmark, the
original WiredTiger needs 25.5MB to persist WAL writes.  Under a Zipf
distribution, this amount varies from 2.58MB to 5.15MB depending on the
skewness of the distribution.  Under a constant distribution with 200B
writes (respectively, 800B), this amount is 10MB (respectively, 40MB).

A naive padding solution which pads all writes to the full slot size
would always require 640MB, regardless of the write-size distribution.
This is equivalent to the 16x best-case or 248x worst-case relative
space overhead.

Figure~\ref{fig:relativeCost} shows the relative space overhead
of \oursys.   This overhead is lowest under the uniform write-size
distribution and when writes have large, constant size.  When the
distribution is highly skewed towards small WAL writes (Zipf with
$\alpha=1.3$), the space overhead is highest because small writes require
a lot of padding.  For a given distribution, larger segment sizes always
result in larger space overhead.

\subsection{Recovery time}

For this experiment, we deliberately crash the system between checkpoints
at times corresponding to different WAL sizes.

\begin{figure}[!h]
    \centering
    \includegraphics[width=0.4\textwidth]{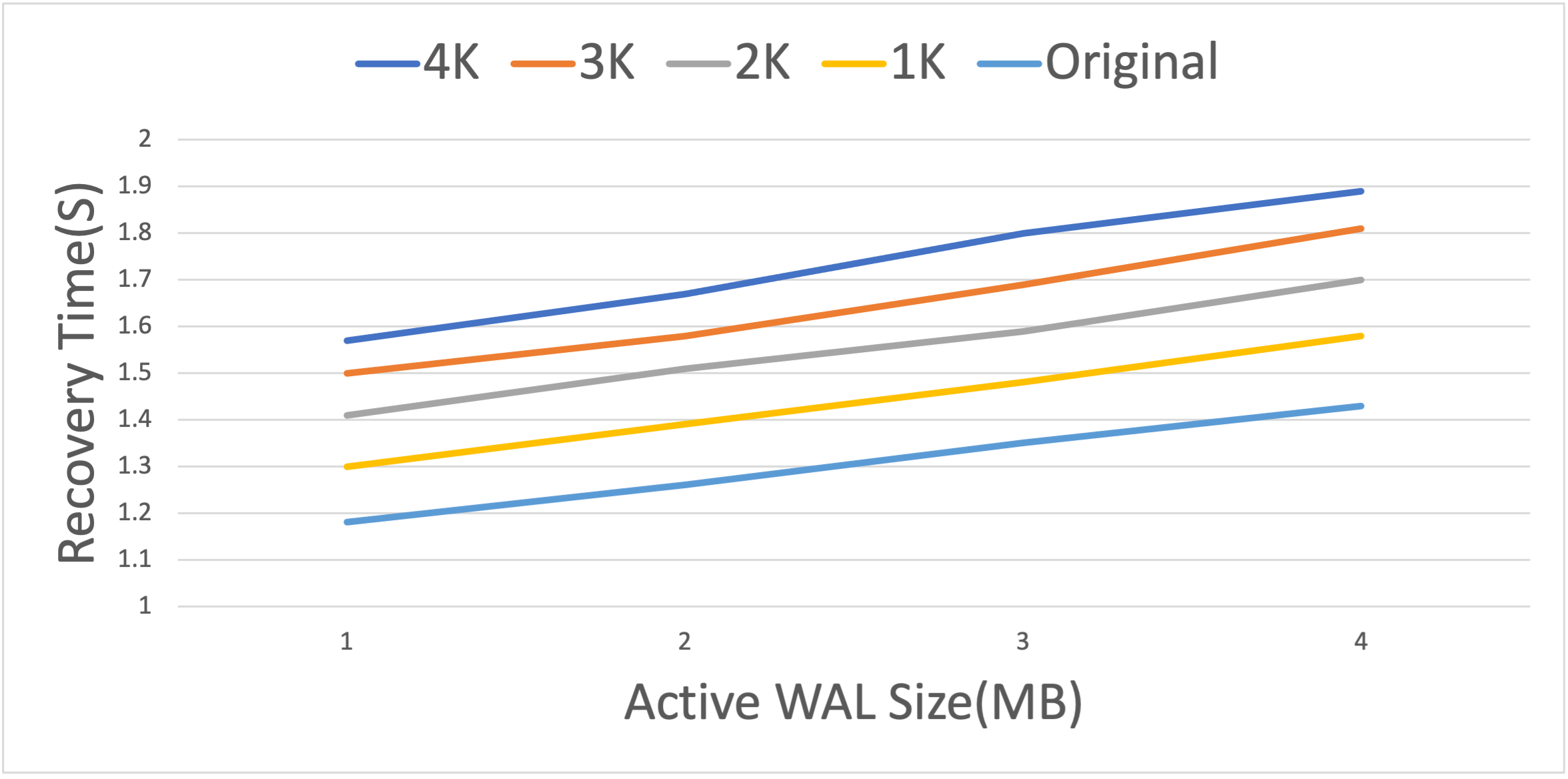}
    \caption{Recovery time.}
    \label{fig:recoverytest}
\end{figure}

Figure~\ref{fig:recoverytest} shows that the original WiredTiger needs
between 1.18s and 1.43s to recover for all WAL sizes.  \oursys increases
recovery time by $10-30\%$ because it requires more I/O when reading WALs
from disk during recovery.

%% file: sections/Limitation.tex
\oursys aims to eliminate variable-sized WAL writes, which are the main
source of fine-grained leakage to compromised storage in databases and
data warehousing systems that separate query processing and storage.
Eliminating variable-sized writes is important because, as we showed
in Section~\ref{sec:attack}, they reveal information even when
encryption-at-rest is deployed.

\oursys does not address coarse-grained leakage from the size of
checkpoints.  Checkpoints typically incorporate a large number of
individual writes, and we are not aware of any method for inferring
sensitive information from their sizes.  Possible less-sensitive
inferences include inferring the specific application operating on MongoDB
(for the purposes of this paper, we assumed that the application is
known), its total memory footprint, and a rough estimate of its write
activity.


\oursys does not address information leakage due to the relative timing
of segment writes.  A storage adversary might be able to infer that two
or more WAL writes in quick succession are part of the same application
write, and use this information to estimate the size of the latter.
Inference attacks described in Section~\ref{sec:attack}, however, require
access to the exact sizes of WAL writes.  If these writes are available
only at segment granularity (i.e., in 1K or larger increments), accuracy
of adversarial inference is significantly reduced.  Furthermore, we are
not aware of any method to distinguish, say, two segment writes resulting
from a slot filled by a large application write and two identically sized
and timed segment writes resulting from a slot filled by multiple small,
concurrent writes.

%% file: sections/newRelated.tex
Prior work on information leakage from encrypted
storage~\cite{cryptoeprint:2016:895,cash2015leakage,islam2012access}
focused on \emph{leakage-abuse attacks} against ``leaky'' encryption
schemes that permit keyword searches, comparisons, and other queries
on encrypted data.  In this paper, we focus on conventional at-rest
encryption that does not permit any queries on the encrypted data and
investigate leakage from the \emph{sizes} of encrypted writes performed
by the query processor.

There has also been work on leakage from access
patterns~\cite{8835292,cryptoeprint:2019:011} and hiding these patterns
by smoothing write frequencies~\cite{grubbs2020pancake}.  Neither these
attacks, nor defenses deal with leakage from the sizes of individual
writes, which is the topic of this paper.